\documentclass{article}

\usepackage{arxiv}

\usepackage[utf8]{inputenc} % allow utf-8 input
\usepackage[T1]{fontenc}    % use 8-bit T1 fonts
\usepackage{hyperref}       % hyperlinks
\usepackage{url}            % simple URL typesetting
\usepackage{booktabs}       % professional-quality tables
\usepackage{amsfonts}       % blackboard math symbols
\usepackage {color}
\usepackage{graphicx}

\title{An Extensive Review of Computational Dance Automation Techniques and Applications}

\author{
  Manish Joshi \\
  School of Computer Sciences\\
  KBC North Maharashtra University\\
  Jalgaon, MS, India, 425001 \\
  \texttt{joshmanish@gmail.com} \\
   \And
 Sangeeta Jadhav \\
  S.S.Dempo College of Commerce and Economics\\
  Altinho, Goa, India. \\
  \texttt{sangeetafromgoa@gmail.com} \\
}

\begin{document}
\maketitle

\begin{abstract}
%\lipsum[1]
Dance is an art and when technology meets this kind of art, it's a novel attempt in itself. Several researchers have attempted to automate several aspects of 
dance, right from dance notation to choreography.  Furthermore, we have encountered several applications of dance automation like e-learning, heritage preservation, etc.  Despite several attempts by researchers for more than two decades in various styles of dance all round the world, we found a review paper that portrays the research status in this area dating to 1990 \cite{politis1990computers}. Hence, we decide to come up with a comprehensive review article that showcases several aspects of dance automation.

This paper is an attempt to review research work reported in the literature, categorize and group all research work completed so far in the field of automating dance. We have explicitly identified six major categories corresponding to the use of computers in dance automation namely dance representation, dance capturing, 
dance semantics, dance generation, dance processing approaches and applications of dance automation systems. We classified several research papers under these categories 
according to their research approach and functionality.  With the help of proposed categories and subcategories one can easily determine the state of research and 
the new avenues left for exploration in the field of dance automation.
\end{abstract}

% keywords can be removed

\section{Introduction}
%\label{}
Efforts of combining dance and computational power can be traced back to 1967. Being a domain that needs relatively more innovation and creativity than mere 
following standard procedures, dance was the slowest to adopt technology. The earliest attempt was published by A. Michael Noll \cite{NollMichael} in Dance Magazine in 1967, 
although New York based choreographer Merce Cunningham also did the same \cite{Cunningham}. 

In 1978, Savage et al. \cite{savage1978} presented an interactive computer model Choreo. Thereafter several research papers were presented that proposed and 
demonstrated the use of computers to refine as well as excel various 
aspects of dance with the 
advancements in technology.  We have enlisted several dance aspects that are being worked upon. It includes dance representation, 
dance capturing, 
dance interpretation, dance generation etc. In this review paper we have presented details of the research work proposed for enhancement of various 
aspects of dance by 
various researchers. Our objective is to provide a comprehensive analysis of the research issues and efforts in the area of computer assisted 
dance automation. 
We propose to use a term \textit{'dance informatics'} to refer research work in the area of computer assisted dance automation.

We have discussed several aspects of dance informatics as follows.  Every dance style has its own notations to represent the dance. 
Researchers have come up with 
dance grammar concept too. We have discussed research work related with dance representation in section 2. The techniques to capture the dance 
and then process it 
further have changed along with time. Several research papers have presented different ideas and proposals to capture dance. All these 
research contributions are 
summarized in section 3. Section 4 presents research experiments that associate semantics with dance. Various approaches of dance 
interpretation using annotation, 
ontology etc. are presented in this section. Computer aided choreography is a keyword that corresponds to automation of the dance generation process.The features of Computer aided choreography are mentioned 
in greater details 
in section 5. Several approaches including probabilistic model, evolutionary model, classification, multi agent systems etc. are   proposed by 
researchers to automate and 
process dance. A list of such different approaches used in dance informatics to enhance the dance is presented in section 6.  Some recent efforts 
of dance 
visualization and dance robotics are also discussed separately in section 7 and section 8 respectively. 
The dance informatics research has certain real life applications. A list of such applications and use of dance informatics in specified applications are
described in section 9, followed by conclusions in section 10.

\section{Dance Representation}

The foremost and essential step in dance automation is dance Representation. In order to comprehend and 
communicate dance steps correctly, a few dance representation techniques are already exist.

\begin{figure*}
	\begin{center}
		\includegraphics{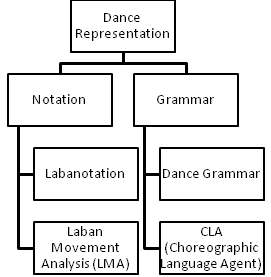}
	\end{center}
	\caption{Dance Representation} 
	\label{DanceRepresentCategory}
\end{figure*}

Just like music notations, which are considered as standards for storing and 
playing it across various 
platforms, it is essential to represent dance standards. Various methods are used to represent dance. Computational power is utilized to 
represent dance by 
developing systems that can work around dance notations. Dance representation can be accomplished using dance notations as well as using dance 
grammar. Figure \ref{DanceRepresentCategory} outlines both of these aspects of dance representation and the research related to these aspects 
is described in subsequent sections.  

\subsection{Notation}

The primary use of Dance notations is the preservation of classical dance documentation. It can be further used for analysis and reconstruction 
of choreography and 
dance forms or technical exercises. Many different forms of dance notations have been created but the two main systems used in Western culture 
are Labanotation (also known as Kinetography Laban) and Benesh Movement Notation \cite{wiki}. Eshkol-Wachman Movement Notation and DanceWriting are also in use, but to a lesser extent \cite{wiki}. 
%(http://en.wikipedia.org/wiki/Dance_notation)
A lot of research work can be noticed dedicated especially in the area of dance notation and its standardization especially, Labanotation 
followed by Latin 
American dance styles and then Indian Classical dance. Latin American dance styles like Ballroom, Foxtrot, Waltz has been used successfully 
for several research 
attempts while Indian classical dances like BharataNatyam, Kutchipudi, Oddisi, etc. have relatively fewer research attempts. 

%
%A Notation allows objective documentation of dance (similar to music notation). A number of Dance notations exist for western dances, the most 
%common being 
%Labanotation and Benesh notation. Labanotation is the most common notation in North America. 
%
Dance Notation Bureau in New York has been 
working for more than six 
decades to disseminate dance scores recorded with this system \cite{DanceNotations}. The resulting archive provides scholars, dancers, 
students, performers and 
the public with an easily accessible detailed record of choreography that allows dances to be studied.

%Automating dance choreography and capturing the movements have been done previously. Dance Representation can be done using existing notation. 

\subsubsection{Labanotation}

A noted Russian BharataNatyam dancer, Annemette Karpen in her paper \cite{annemette_p_karpen_labanotation_1990} has tried to solve the problem 
of notation for BharataNatyam by using Labanotation. She 
has stated that video recordings are not a good method of presenting dance movements since camera catches only one angle that too from the 
eyes of the director 
and not the finest details. She also stated that Sutton Dance writing is an easy method for notating dance but available only for western 
dances and so she tried 
using the same for BharataNatyam and realized its limitations in not being able to capture the finer movements of BharataNatyam's facial 
expressions, hand 
gestures, neck, shoulder movements and intricate footwork. So she also tried the same using Labanotation which is again used for Western 
dances only and had 
similar problems as above. Ebenreuter \cite{ebenreuter_transference_2006} has attempted to design an interface to facilitate the exact documentation of dance notation while the LabanDancer system 
developed by Wilke et al. \cite{wilke_dance_2005} helped to translate the recorded Labanotation scores into 3-D human figure animations.
There are few dancers or choreographers who can read notation or even capable of producing scores. Thus a computer based tool to animate the 
notation would prove 
very helpful. Labanwriter is used for creating and editing dance scores. LifeForms \cite{lifeForms} is a product of Credo 
Interactive for 3D choreography, animation and motion picture which helps to experiment with patterns of movement in animated human figure for Labanotation.

\subsubsection{Laban Movement Analysis (LMA)}

LMA provides a model for observation, description and notational system for human movements. Implementation of the same in a computer has been done through 
Bayesian approach \cite{j_rett_bayesian_2008} and also by 3DTI in which choreographic process was enhanced by Nahrstedt et al. \cite{klara_nahrstedt_computational_2007} using a 3D tele-immersive (3DTI) room surrounded by on 
multiple 3D digital camera and displays from a remotely placed dancer.

\subsection{Grammar}
Dance can be represented by capturing underlying rules of all possible movements. A grammar of all possible movements can be formulated and it shall help in further representation and processing. Some of the 
approaches use grammar to store body movements or all possible dance rules. Following subsections describe two such grammar related approaches. 

\subsubsection{Dance Grammar}

Dance grammar is used for representation of dance (as shown in Figure 1) as well as for dance  interpretation (as shown in Figure 3).

A grammar for gestures captured on a multi-touch device is derived and used by Puig et al. \cite{puig_fingers_2010} to identify the different choreographic moves which shall help in recalling all 
corresponding clips from the movie.  Bradford et al. \cite{bradford_application_1995} developed a tool CorX for exploration of dance grammar which is a set of executable if-then rules. Bradley et al. \cite{bradleyLearning1998} have given a corpus based grammar of joint movements. By using simple machine learning and search techniques authors have claimed to automate the process of 
animation for a prescribed body posture. Bull \cite{bull_formal_1996} produced a formalised grammar of Aerobics Dance Exercise. This formed the Aerobics Corpus which was collected from a National Survey. Using Computational Linguistic techniques, he extracted a formalised grammar for Aerobic Dance choreographic moves. 

\subsubsection{Choreographic Language Agent (CLA)}

Choreographic Language Agent (CLA) \cite{delahunta_future_2009} helped to bridge the gap between the notations, sketches, diagrams and text done by the choreographer on a notebook and his 
thinking process. Thus an unique method was used to augment the thinking process of the choreographer.

\section{Dance Capturing}

Various techniques have been used for capturing dance movements of different styles. The most common being motion capture techniques and use of sensors. 
Dance motion can be captured using multi view vision camera or monocular vision camera. The obvious difference between usages of both is the additional cost 
factor and also the various dimensions of the object.

Capturing and modeling of all the dance movements correctly results in efficient processing too. Several attempts have been made by researchers to capture dance mainly with the help of sensors and motion capture cameras (Figure 2). The details of techniques used and research work thereof is presented in following subsections.

\begin{figure*}
	\begin{center}
		\includegraphics{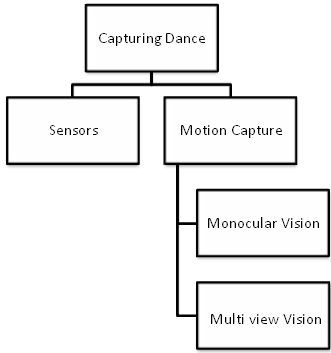}
	\end{center}
	\caption{Dance Capturing} 
	\label{DanceCaptureCategory}
\end{figure*}

\subsection{Sensors}

A device which receives and responds to a signal on a certain preset event (say 
`when touched') is a sensor. Several experiments have been carried out 
with the help of dance experts who wear different sensors on their body. The sensors are monitored to obtain dance movement data.

The touch interactive robot trained by Curtis et al. \cite{curtis_dance_2011} has touch sensors residing on its tail, legs, back and head along with contact switches on the bottom of 
each foot which was taught motions via demonstration and reinforcement. Using 41 markers on dancers Qian et al. \cite{qian_gesture-driven_2004} have developed a real time gesture driven 
interactive system. This marker based motion capture systems were used to provide real-time marker positions in global space. Paradiso et al.  \cite{paradiso_design_2000} have used a 
sensor system along with a small microprocessor and wireless transmitter in a pair of dancing shoes to capture the various degrees of freedom for the feet which 
can be made applicable for a series of computer-augmented dance performance. Aylward et al. \cite{aylward_sensemble_2006} have used a compact , wearable sensor system which enabled real 
time collective activity tracking for interactive dance. The dancer wore a wireless sensor at the wrists and ankles. Takasi et al. \cite{takahashi_dance_2009} used a  Motion Capture system called as Eva 
Hires which is an optical motion capture system introduced in the Hiroshima city university for synthesizing dance data automatically for a user given music, even if the user has no knowledge of music at all. A multimedia system  \cite{barry_motion_2005} uses a novel motion classification scheme. This is a wearable computer which aims to achieve  a 3-D dance style classification for  Butoh dance. This is a  contemporary dance improvising method originating in Japan.  

\subsection{Motion capture}
Motion capture is defined as ``The creation of a 3D representation of a live performance'', by Alberto Menache in the book Understanding Motion Capture for Computer Animation and Video Games. Thus Motion Capture technology shall help in detailed analysis of the captured dance movement and use that data to develop ways to analyze the movement in detail.

\subsubsection{Monocular Vision}

Chen et al. \cite{chen_markerless_2005} have used a single camera without any markers for Motion Capture to obtain 3D motion parameters of a human figure. This tracking of human figure on video has used image silhouettes. Paul et al. \cite{paul_virtual_1998} have used a single camera to metamorphose the user into a virtual person who can be an off-site coach using 
low band-width joint motion data to permit real time animation.  This metamorphosis involved altering the appearance of the person in this case a kathakali dancer 
since this dance style has elaborate costume and make-up which is very time consuming.  The user did not have to wear any hardware device and the paper aimed at 
making gesture tracking simpler, cheaper and user-friendly. Curtis et al. \cite{curtis_dance_2011} have trained a robot named Pleo from Innvo Corporation for treating autistic 
children with dance therapy. In this, a coloured camera is used in the nose of the friendly dinosaur looking robot, Pleo. Using a single camera for motion capture 
of BharataNatyam \cite{mamania_2004}, a semiautomatic method was developed to track the body parts using skin colour detection. The authors restricted their model up 
to upper body part only.

\subsubsection{Multi view Vision}

Lapointe et al. \cite{lapointe_choreogenetics} have used genetic algorithm for real time generation of human computer choreography through choreogenetics algorithm. Using motion capture 
with 8 cameras and 8 PCs, Nakazawa et al. \cite{nakazawa_digital_2002} have used motion analysis method for recognising the structure of human dance motion and motion primitives for a 
Japanese folk dance ``Soran Bushi''. Concatenating these primitives they have generated new dance motions using inverse kinematics and dynamic balancing 
techniques. 

Brand et al. \cite{brand_style_2000} have generated stylistic motion sequences from motion patterns through motion capture sequences. Data was captured by physical markers placed on 
human actors, over short interval of time, in motion capture studios. Qian et al. \cite{qian_gesture-driven_2004} have tackled motion capture's marker occlusion problem by developing a 
real-time marker cleaning algorithm. Using 8 camera VICON systems for training, the multimodal feedback engine produced visual and audio feedback to the performer.

\section{Dance Semantics}

\begin{figure*}
	\begin{center}
		\includegraphics{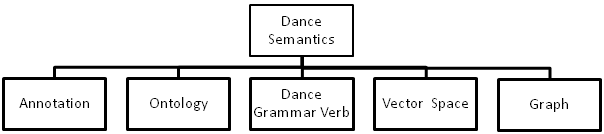}
	\end{center}
	\caption{Dance Semantics} 
	\label{DanceSemanticsCategory}
\end{figure*}

After representing and capturing dance, computational power can be used to process it. The model used to represent the dance, must enable user to interpret the processed dance steps. Thus processing the stored dance steps shall help in interpreting the 
meaning of the particular style of dance. Several techniques have been applied to understand the semantics of dance data from annotation, ontology, dance grammar 
or dance verbs, graph to vector space. These techniques used to interpret dance semantics 
have helped in the process of automating choreography. 
Figure \ref{DanceSemanticsCategory} shows techniques under dance semantics and the details of each technique follows in subsequent subsections.

\subsection{Annotation}
Puig et al. \cite{puig_fingers_2010} have experimented with Thierry de Mey's Project for gestural annotation of dance video recordings. The grammar used by Mey to identify the 
different choreographic moves within the show and the annotation was processed through IRI's software. To each identified gesture of the dancers, corresponding 
moves of the observer's two hands were recorded through a multi-touch sensitive surface. Cabral et al. \cite{cabral_multimodal_2011} have designed a video annotator for tablet PC using 
touch input interfaces. This is used for contemporary dance as a creative tool by choreographer to improve his work during rehearsal, live performances for later 
review or sharing notes with performers which can in future be also used for general web-based archive. Mallik et al. \cite{mallik_nrityakosha} have automatically annotated new 
instances of digital heritage constructed ontology for Indian Classical Dance BharataNatyam  and Odissi to train multimedia data. E-dance project \cite{helen_bailey_dancing_2009} showed how grid-based hypermedia and semantic annotations were used for capturing and rendering the choreographic practices. For content based multimedia access, concept 
recognition using ontology is used by Malik et al. \cite{mallik_preservation_2010}. The automatically annotated new instances enabled creation of semantic navigation environment in a 
cultural heritage repository. Video annotation by Malik et al. \cite{mallik_using_2009} using the power of MOWL has provided an effective video browsing interface to the user through 
a Bayesian Network for Indian Classical Dance such as BharataNatyam, Odissi, Kutchipudi and Kathak including music performances like Hindustani and Carnatic 
music. 200 videos of about 10 to 15 minutes duration were used of all these Indian Classical Dances for experimentation.

\subsection{Ontology}

Mallik et al. \cite{mallik_preservation_2010}, \cite{mallik_nrityakosha} have constructed ontology for Indian Classical Dance BharataNatyam and Odissi to train multimedia data and automatically annotate new 
instances of digital heritage. The domain knowledge has been encoded in ontology and has provided methods to co-relate this to the audio-visual recordings and 
other digital artifacts. An ontological framework for Indian Classical dance by Malik et al. \cite{mallik_using_2009} offered a robust ground for several multimedia search, 
retrieval and browsing applications. The system was self enhancing where ontology was refined from annotated data and data annotation was improved based on fresh, 
refined knowledge from the ontology.

\subsection{Dance Grammar (Verbs)} 

A  gestural annotation of dance video recordings to codify Mey's \cite{puig_fingers_2010} own movies and musical pieces in which he has elaborated a grammar of gestures. A multi touch 
sensitive surface records the observers' two hands for each identified gesture which can help identifying all similar clips in the movie. Hsieh et al. \cite{hsieh_generating_2005} have 
used Newton's law for presenting a set of dynamic models according to some dance verbs for contemporary dance. This paper was done with an aim to assist in 
computer-aided choreography and overcome the difficulty of dynamic based animation. Bradford et al. \cite{bradford_application_1995} has described a program that uses Artificial Intelligence 
for dance. An if-then rule driver approach was used which described motion to create a sequence of dance phrases and this was used to generate choreography for 
multiple dances. A complex set of rules are thus created, executed and evaluated for patterns of dance.

\subsection{Vector space}

Jadhav et al. \cite{sangeeta_jadhav_modelling_2012} have used vector space for modeling BharataNatyam. All major limbs of the body; head, hands (both right and left), waist and legs (both right 
and left) are coded as per the orientations, positions and strict norms of the Indian Classical dance, BharataNatyam. The major limbs of the body have Sanskrit 
names as per the ancient dance scripture Natyasastra and they have been modeled as per these names and the X, Y and Z axes respectively.

\subsection{Graph} 

A graph based algorithm was proposed for reconstructing 3D model for BharataNatyam dance from tracked data.  
The authors Mamania et al. \cite{mamania_2004} proposed to visualize a pose as a node and a transition between two poses as an edge between corresponding nodes.Sugathan et al. \cite{sugathan_attributed_2014} has extracted features from an attribute relation graph for the upper body poses of basic Bharatanatyam steps which can be useful for classification and annotation of dance poses.

\section{Dance generation}

\begin{figure*}
	\begin{center}
		\includegraphics{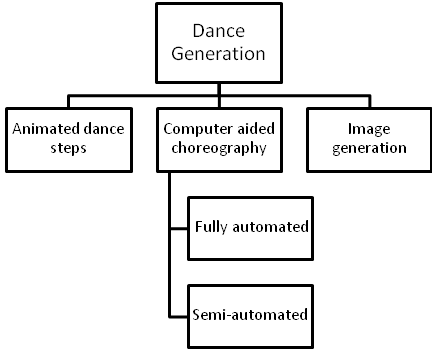}
	\end{center}
	\caption{Dance Generation} 
	\label{DanceGenerationCategory}
\end{figure*}

Dance generation using computer system has been attempted by several researchers and some have been successful in fully or partially automating the process. We enlist a few methods that automated dance choreography by generating dance steps or poses through 3D, using robot motion to generate images. We have identified three major aspects of dance generation namely, animated dance steps, computer 
aided choreography and Image generation as shown in Figure \ref{DanceGenerationCategory}. The details of each aspect are elaborated in three subsections. 

\subsection{Animated dance steps}

Nagata et al. \cite{nagata_analysis_2004} obtained Latin Dance movements using Motion Capture and attempted to extract the difference in the character of movement of  experienced people (Latin people) as well as inexperienced (Japanese people) for Latin dance. Thus after the extraction of natural dance movements, animation was carried out to make use of the outcome. Using advanced 2D research techniques, they have confirmed a phase difference in the movement of shoulders and hips considered to be a characteristic movement of 
experienced Latin dancers. They claim that Japanese people are unfamiliar to a motion called  ''isolation''.  Wilke et al. \cite{wilke_dance_2005} have developed the LabanDancer to translate Labanotation dance scores into 3-D human figure animation 
for a wide variety of different movements. The main reason being fewer dancers can read notation and even fewer are capable of producing scores. This can be a 
teaching tool for dance choreographers and students. Animation was done using Life Forms software by Sukel et al. \cite{sukel_presenting_2003} and all the participants performed three 
formal Ballet movements in same order and they filled out a demographic sheet The authors felt that computer animation provided a better joint segmentation thus 
improving learning as compared to videotapes. Mazumdar et al. \cite{rwitajit_majumdar_framework} have generated a library of body movements based on rules of BharataNatyam and human body 
constraints which are further used to generate the dance steps for pure dance movements.

\subsection{Computer aided choreography(CAG)}

CAG is further sub categorized into fully automated and semi automated. CAG as discussed in following subsections.

\subsubsection{Fully automated}

Hagendoorn has transformed a dance studio \cite{hagendoorn_emergent_2002} into a laboratory for studying complex systems for dance choreography. He has tried to find different equivalence 
relations and classes as they apply to dance so that fascinating patterns emerge within a group of dancers. He has formulated a set of what he called as nature 
inspired rules that determined the interaction among a group of dancers. The system results in patterns that indicate how dancers should move on stage. Although the 
focus of the work is limited to movement of dancers on stage, the system generates fully automated choreography. Nakazawa et al. \cite{nakazawa_dancing_2009} have used genetic 
algorithms for a fully automated system of waltz choreography. They used mutation and crossover for exploring possible solutions to obtain a global optimum. The system generated satisfactory results by using majority of the 
stage, keeping partners facing each other and dancers on stage. The authors claimed that the system resulted within 10 \% of the optimal choreography.  Using the Choreogenetics algorithm \cite{lapointe_choreogenetics}, choreographic variants were obtained for five basic 
movements. These were selected based on aesthetic criteria. Lapointe et al. \cite{lapointe_dancing_2005} have shown that the best mutants closely match with the virtual dancer and thus the duets generated by the algorithm are not entirely random. Takahashi et al. \cite{takahashi_dance_2009} proposed a dance synthesis system that utilizes the motion capture data with Computer Graphics software Maya according to the impressions of music. Yu et al. \cite{yu_tour_2003} tried to demonstrate the workability and usefulness 
of computer generated choreography by using swarm toolkit and Life forms software with multi-agent system.  The 
swarm toolkit was used to generate a sequence of dance steps which was later animated and expert evaluation is 
sought. Jadhav et al. \cite{sangeeta_jadhav_manish_joshi_jyoti_pawar_art_2012},\cite{IJCSA_2014},
\cite{Taylor_Francis_2015},\cite{COMAD_2013} have created an Indian Classical Dance choreography 
generator termed as ArttoSMart which displays several possible sequences for pure dance movements in
BharataNatyam, given the number of beat and a starting pose as an input. Thus the user can choose the best 
possible sequence from the available options.

\subsubsection{Semi-automated}

Stuart et al. \cite{bradleyLearning1998} have developed and used corpora of human movements comprising of ten Balanchine ballets to select a movement sequence that would naturally occur between a 
given pair of body postures. They have applied techniques from graph theory, Artificial Intelligence and statistics to the above corpus of movement sequence. Interpolation methods are described in this paper to automatically construct interpolation sequence that suggest moves from one specified body posture 
to another in a physically and stylistically coherent fashion. Curtis et al. \cite{curtis_dance_2011} has designed a system that could be trained to learn dance movements through 
visual and haptic cues. With human assistance the robot could learn and perform dance steps. Bradford et al. \cite{bradford_application_1995} utilized rule driver embodying a heuristic 
algorithm for choreography of dance. They have designed a system CorX which may prove valuable to choreographers as an aid to the creative process but many of the 
details of interpretation are left to the human choreographers.

\subsubsection{Image generation}

Pattanaik \cite{pattanaik_stylised} has tried animating a few BharataNatyam karanas using stick figure model initially and finally a stylised volumetric model was used which could convey the position of the body efficiently and correctly. Jadhav et. al \cite{Stickfigure_2014} have automated the process of 2D Stick Figure generation from the 30 attribute human body model \cite{sangeeta_jadhav_modelling_2012}. 

\section{Dance Processing Approaches}

\begin{figure*}
	\begin{center}
		\includegraphics[angle=90]{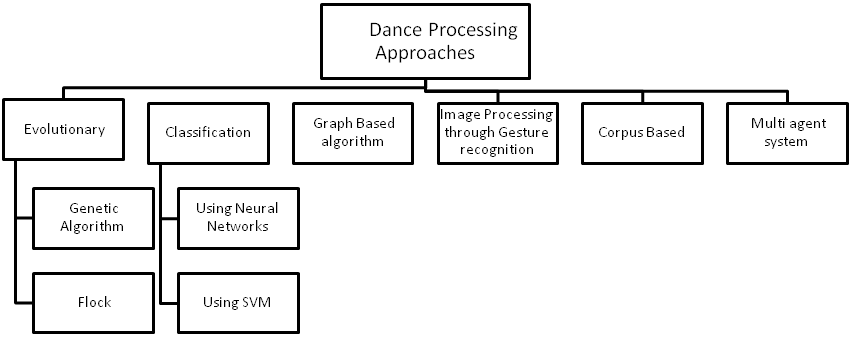}
	\end{center}
	\caption{Dance Processing Approaches} 
	\label{DanceProcess}
\end{figure*}

Capturing and modeling of all the dance poses effectively ensures efficient processing to obtain automated dance movements. Several attempts have been made by researchers in various ways 
to process these captured and modeled data, refer Figure \ref{DanceProcess}. These approaches and corresponding research work are explained in later subsections. The approaches discussed in this paper are as follows.  
Evolutionary Programming using Genetic, Flock and Ant optimization Algorithms; Classification using Neural networks and Support Vector Machines; Image Processing 
for gesture recognition; Graph based algorithms; Corpus Based and Multi agent system.

\subsection{Evolutionary Approach}
This approach works on the powerful principle of evolution i.e. survival of the fittest \cite{dasgupta_dipankar_evolutionary_1997} which models natural phenomena like genetic inheritance 
and Darwinian strife for survival using heuristic search.

\subsubsection{Genetic Algorithms}       
A Genetic Algorithm based paper \cite{lapointe_choreogenetics} that generates aesthetic choreography through mutations and selection and a new 
algorithm is applied to simulate the 
evolution of a sequence of dance movements. Using GA to create human computer choreography for real-time performance environments, Lapointe et 
al. \cite{lapointe_dancing_2005}  have 
created human computer duet by using motion capture technique on actual performers and coded the same with LIFE animation software to create a 
virtual vocabulary 
of four movements: run, jump, turn and fall.  Nakazawa et al. \cite{nakazawa_dancing_2009} have used genetic algorithms for a fully automated 
system of waltz choreography. The fitness function was designed with respect to following equally weighted factors like the couple's position, use of stage, step sequence, closeness to ideal steps, measurement of stage and so on. 
%
%The use of 
%mutation for exploring the possible solutions to a global optimum keeping a large population size and low mutation rate is noticed here. The 
%system generated 
%satisfactory results within 10\% of the optimal choreography by using majority of the stage, keeping partners facing each other and dancers on 
%stage. 
%
Jadhav et al. \cite{sangeeta_jadhav_manish_joshi_jyoti_pawar_art_2012}, \cite{IJCSA_2014}, \cite{Taylor_Francis_2015}, \cite{COMAD_2013} have used Genetic Algorithm to find static dance poses for 
single beat and Multi Beat with the use of a fitness function which tries to converge with results neither too close nor far from the ideal pure dance movements called as ''adavus'' for the  Indian Classical Dance, BharataNatyam.

\subsubsection{Flock}

Hagendoorn [31] has used flock technique for generation of enticing patterns of dance. He says that within a flock only nearest neighbors are visible and hence 
it is used for self- reinforcing of dance patterns. Rules governing the behavioral pattern of agents are listed out and most of the rules are inspired by the 
flock of birds or swarm behavior.

\subsection{Classification}
Classification is a technique where the user knows ahead how classes are defined. It is necessary that each record in the data-set, used to build the classifier. already have a value for the attribute used to define classes. Dance Processing can also be done using various classification techniques like Neural Networks and Support Vector Machines.

%Because each record has a value for the attribute used to define the classes, and because the 
%end-user decides on the attribute to use, classification is much less exploratory than clustering. The objective of a classifier is not to explore the data to 
%discover interesting segments, but rather to decide how new records should be classified. Thus it helps in predicting group membership for data instances. [cite 
%paper] Thair Nu Phyu , Proceedings of the International MultiConference of Engineers and Computer Scientists 2009 Vol I ,IMECS 2009, March 18 - 20, 2009, Hong 
%Kong Survey of Classification Techniques in Data Mining

\subsubsection{Neural Networks}
Dubbin et al. \cite{dubbin_learning_2010} presented a program that takes advantage of interactive evolutionary computation and Artificial Neural Networks to train virtual humans to 
learn to dance. The dancers were controlled by ANN. They have efficiently solved the problem to parse sound in a way that ANN could interpret it.

\subsubsection{Support Vector Machine(SVM)}
Using k-means clustering algorithm Mallik et al. \cite{mallik_preservation_2010} have trained an SVM classifier for classifying the media patterns. A training set of video segments were labeled by domain experts which helped in creating the multimedia enriched ontology and use of machine Learning algorithms re validated the same with the use of low level media features and SVM classification.  This was further used to interpret the media features extracted from a larger collection of videos to classify them into semantic groups. Sharma \cite{sharma_Apratim_2013} has used an action classifier for the basic BharataNatyam dance moves called as adavus using SVM.

\subsection{Graph Based algorithms} 
Bradley et al. \cite{bradleyLearning1998} have designed interpolation algorithm for a Ballet dancer's body postures. These movements remain consistent from one 
prescribed form to another. The use of transition graphs for capturing transition probabilities for every body-joint and generating a 
small corpus for ballet sequences and finally interpolation sequences and applying A* search has resulted in learning the grammar of 
dance. The graph-theoretic methods learn the grammar of joint movements in a given corpus.

%A multimedia system  \cite{barry_motion_2005} which aims to 
%achieve using body-worn acceleration sensors and a wearable computer leading to a 3-D dance style classification scheme for a 
%contemporary dance improvising method originating in Japan called as Butoh dance.  

%Brand et al. \cite{brand_style_2000} have captured data through physical 
%markers placed on human actors over short intervals in motion capture studios. Use of Machine Learning algorithms synthesized novel 
%motion data into more graceful dance of an expert. This model was used to generate choreography and synthesize virtual motion capture in 
%many styles.

Feature extraction was done by Hariharan et al. \cite{hariharan_recognizing_2011} by generating a feature vector  for distinguishing between different 
gestures of a BharataNatyam dancer's single hand gestures. The silhouette was extracted followed by the generation of the 
corresponding skeleton and the evaluation of the gradients at its end points. Morphological operators are used to obtain a skeleton which corresponds to graph called as connectivity graph. Several such connectivity graphs for different hand gestures are shown. Sugathan et al.\cite{sugathan_attributed_2014} has proposed a graph based model for identifying and classifying the 2D poses of a BharataNatyam dancer's upper body.

\subsection{Image Processing through Gesture recognition}
Quin et al. \cite{qian_gesture-driven_2004} have proposed a gesture recognition engine which provides real time recognition of the performer's gesture, based on the 
3D marker co-ordinates. Hariharan et al. \cite{hariharan_recognizing_2011} developed a prototype for the recognition of the 28 single hand gestures of BharataNatyam called as Asamyukta Hastas in a 2D  space using image processing techniques whereas Saha et al. \cite{saha2014BNhand} have used boundary of the hand gesture and texture based segmentation to sort out the flaws for recognition of the same single hand gestures. Emotions of the Indian Classical Dancer have been captured through the Kinect Camera by Saha et al. \cite{saha2013gesture} for creating a gesture recognition algorithm and in \cite{saha2013fuzzyBN} for automatic BharataNatyam hand gesture recognition. Sharma  \cite{sharma_Apratim_2013} has used the Kinect camera in his M.Tech. thesis, to capture and recognize the basic BharataNatyam adavus.

\subsection{Corpus Based}

Bradley et al. \cite{bradleyLearning1998} have developed a corpus- based interpolation algorithm for movement sequences to achieve a ballet dancer's move from one body posture to another for computer animation. Bull \cite{bull_formal_1996} has designed a corpus of Aerobic Dance Exercise routines for analysis using linguistics techniques. He has named it ACCOLADE: A Computerized Corpus of Legal Aerobic Dance Exercises. Later on he used the real time animation of dancers with the NUDES system developed by University of Sydney. This was successfully applied to his Aerobics project.

\subsection{Multi agent system}

The prototype system proposed by Hagendoorn \cite{hagendoorn_emergent_2002} is based on the concept of modern multiagent system. This term (multiagent system) is not 
explicitly used by the author but each dancer is assumed to be an agent and determines his/her behavior by sensing the state of other 
agents (individual or as a group). 

\section{Dance Visualization}
Live dance performances can use different forms of multimedia. These forms can be preplanned animation sequences or the use of digital video with appropriate sound at the background along with live dancers. These performers can interact with this pre programmed display. A more technically challenging system would be to sense the live dancers and modify the sound or images accordingly or having two distant locations for rehearsal and performance. Thus various possibilities are available and a challenging problem of computer graphics and interactive technologies for live performers can be observed here.

Stage visualization has been effectively done by Calvert et al. for enhancing live performances \cite{calvert2005applications} and edited by Potel. The use of computer graphics has been conceptualized to visualize 
choreography, composing, editing and animating dance notation for enhancing live performances. This tool helped the choreographer to try 
out ideas before meeting with the live dancers. The software has human male and female figures for ballet and modern dance and thus a 
choreographer can easily plan even in scarce dance studios. A virtual dance studio has been created by National Arts Center, Canada 
through their website \cite{Artsalive} where various movements of a ballet dancer are placed and options available for the viewer to choose 
the start, middle and end sequence. Finally these entire movements are co-ordinated in a sequence and played for visualization purpose. 
Dubbin et al. \cite{dubbin_learning_2010} presented a model called Dance Evolution allowing the user to train virtual humans to dance to MIDI songs or raw audio. This was implemented using Panda 3D simulator.

\section{Dance Robotics (Humanoid)}

Humanoid robots are used in several research areas from domestic help to space and dance is one such area where it has been successfully 
implemented.
Shinozaki et al. \cite{shinozaki_concept_2007} have  proposed a humanoid dance robot system for a hip-hop dance sequence. Based on a professional dancer's basic 
movements,  sixty dance units were extracted and these short dance motions were concatenated for a robot dance system. These resultant 
concatenated units created huge amounts of dance variations for the hip-hop genre. Gruberg et al. \cite{david_gruberg_development_2010} have experimented with the motions of a robot that are co-coordinated 
automatically  to the music beat. They have used a real time music signal thus enabling humanoid robot to dance 
autonomously. Nakazawa et al. \cite{nakazawa_digital_2002} have proposed a dance motion imitation for humanoid robots through visual 
observation. Using a stochastic controller, Angulo et al. \cite{angulo_aibo_2011} 
have tried to generate dance movements adequate to the music rhythm. They have tried to recreate a human-robot interaction system with 
the Aibo robot.

\section{Applications of Automated Dance Systems}
\begin{figure*}
	\begin{center}
		\includegraphics[width = \textwidth]{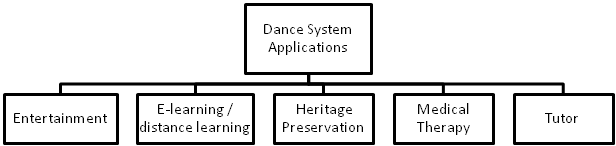}
	\end{center}
	\caption{Dance System Applications} 
	\label{Danceapplication}
\end{figure*}

There are several advantages of computerized modeling of dance. The biggest commercial industry is of entertainment and gaming followed by e-learning, distance learning, computerized dance tutor. Additionally dance systems applications can be found in robots used for medical therapy for differently-abled children and also for the sake of heritage preservation of various classical dance forms and dying art. All these application areas shown in Figure \ref{Danceapplication} and explained thereafter.

\subsection{Entertainment}

Nirvana Technologies developed robot dance system has been used by Shinozaki et al. \cite{shinozaki_concept_2007} for dancing various moves which are short 
concatenated dance motions performed by the humanoid robot. The hip-hop style was chosen over ballet since the later does not have 
specific rules for the details of whole body motions.

\subsection {E-learning / distance learning}

Paul et al. \cite{paul_virtual_1998} showed how a renowned Kathakali dancer from a remote site may provide feedback 
to a student which is an excellent resource for e-learning. Both the master and the disciple can zoom on a particular performance to view 
the scene from a vantage point or repeat action in slower speed. Using a 3D tele-immersive (3DTI) room surrounded by multiple 3D digital 
camera and displays from a remotely placed dancer \cite{klara_nahrstedt_computational_2007} visual stimulations have helped in distance learning. To understand and extend choreographic knowledge, e-dance project \cite{helen_bailey_dancing_2009} was used where performers and spectators were co-present in physical spaces and simultaneously shared multiple, virtual locations. Computer is used as a teacher by Hariharan et al. \cite{hariharan_recognizing_2011} in an e-learning environment to rectify the twenty eight single hand gestures of BharataNatyam (Asamyukta Hastas) performed by students whereas Saha et al. \cite{saha2013fuzzyBN} have been able to improve on time, efficiency and accuracy for the same single hand gestures through the use of fuzzy L Membership function and in \cite{saha2014BNhand} through polygon representation. A mobile based applet was used by Mazumdar et al. \cite{rwitajit_majumdar_framework} to teach BharataNatyam adavus through a puzzle based game. Using the kinect sensor, Saha et al. \cite{saha2013gesture} have extracted positive or negative emotions from the gestures made by the Indian Classical Dancer which can be thus used for learning and evaluating the performance.

\subsection {Heritage Preservation}

Digital Multimedia technology helps in preservation of heritage \cite{mallik_nrityakosha} and also enhances its accessibility over a prolonged period of 
time. This paper has correlated the digital resources with the traditional knowledge of Indian Classical Dance BharataNatyam and Odissi .

A Russian BharataNatyam dancer Annemette Karpen \cite{annemette_p_karpen_labanotation_1990} has used Labanotation to notate BharataNatyam dance so as to record and preserve the heritage of Indian Classical dance.

\subsection {Medical Therapy}
Dance Movement therapy has been used effectively for treatment of wide range of physical and mental disorders. Curtis et al. \cite{curtis_dance_2011} has 
designed Pleo which is a low-cost, off the shelf, robotic platform designed specially to be a trained robotic dance therapy assistant 
capable of engaging autistic children. The desired dance motions and music are provided to the robot by the dance therapist who in turn 
introduces the same to the patient and this can be provided for home use to accelerate the therapy process. Thus a low-cost dancing robot 
has been trained for medical therapy.

\subsection {Tutor}

Sukel et al. \cite{sukel_presenting_2003}  developed a computer based dance tutor which  is a virtual aid to teach formal ballet so that it may provide individualized attention that dance classes lack due to higher student-teacher ratio.

\section {Conclusion} 

Creativity is considered to be a gift and no two persons can be same in this process. A field of art like Dance is considered to be 
entirely a creative process although the basics are very clearly specified and taught in case of Classical Dance forms. Each 
choreographer is known for their particular style and creative form. Using a machine to aid in this creative process has been attempted 
by many well-known creative artists and dance is also a domain known to accept and start experimenting with this. We have reviewed at 
least hundred of such research papers and articles and categorized them accordingly. 

%selected 45 of them to categorise and present the areas of research already 
%ventured and there are many more remaining unexplored. 

We have identified six such major categories for classification and discussed how the research work has been carried out in these areas. 
The Dance forms range from Indian Classical like BharataNatyam, Kutchipudi, etc to Western Classical like Ballet. Also some Japanese folk 
dance form has been used like Soran Bushi and contemporary style like Butoh. Our aims for writing this review of dance papers are to make 
it easier for the researcher who is interested in automation of the choreographic process like us. 
%The resulting choreography shall 
%always be unique and different since a machine is not biased and also it can show many alternatives at the same time that a human brain 
%cannot imagine. 

%% References


\bibliographystyle{elsarticle-num}
\bibliography{InfoSciJor}






% \end{thebibliography}

\end{document}